\documentclass[conference,a4paper]{IEEEtran}
\usepackage[utf8]{inputenc}
\usepackage{times,epsfig,graphicx,epstopdf,listliketab,tabularx}
\graphicspath{{./pics/}}
\DeclareGraphicsExtensions{.pdf,.png,.jpg,.eps}
\usepackage[verbose]{placeins}
\usepackage[cmex10]{amsmath}
\usepackage{balance}
\usepackage{tikz}
\usetikzlibrary{patterns,positioning,datavisualization,arrows}
\usetikzlibrary{datavisualization.formats.functions}
\usepackage{pgfplots}
\pgfplotsset{compat=1.5.1}
\usepackage[format=hang]{subfig}
\usepackage{siunitx}
\usepackage{stfloats}
\usepackage{algorithmic}
\usepackage{paralist}
\usepackage{booktabs}
\usepackage{ifthen}
\usepackage{multirow}
\usepackage{listings}

\def\COPYRIGHTYEAR{2019}	%
\def\CONFERENCE{IEEE International Conference on Communications 2019 (ICC 2019)} %

\def\conferencenotice{Accepted for presentation in: \CONFERENCE}	

\def\copyrightnotice{
	\copyright~\COPYRIGHTYEAR~IEEE. Personal use of this material is permitted. Permission from IEEE must be obtained for all other uses, including reprinting/republishing this material for advertising or promotional purposes, collecting new collected works for resale or redistribution to servers or lists, or reuse of any copyrighted component of this work in other works.
}

\def \notice{
	\begin{tikzpicture}[remember picture, overlay]
	\node[below=5mm of current page.north, text width=20cm,font=\sffamily\footnotesize,align=center] {\conferencenotice};
	\end{tikzpicture}
	\begin{tikzpicture}[remember picture, overlay]
	\node[above=5mm of current page.south, text width=15cm,font=\sffamily\footnotesize] {\copyrightnotice};
	\end{tikzpicture}
}

\usepackage[backend=bibtex,bibstyle=ieee,giveninits=true]{biblatex}
\usepackage[british]{babel}
\usepackage{csquotes}
\DeclareFieldFormat{bibentrysetcount}{\mkbibparens{\mknumalph{#1}}}
\DeclareFieldFormat{labelnumberwidth}{\mkbibbrackets{#1}}

\defbibenvironment{bibliography}
  {\list
     {\printtext[labelnumberwidth]{%
    \printfield{labelprefix}%
    \printfield{labelnumber}}}
     {\setlength{\labelwidth}{\labelnumberwidth}%
      \setlength{\leftmargin}{\labelwidth}%
      \setlength{\labelsep}{\biblabelsep}%
      \addtolength{\leftmargin}{\labelsep}%
      \setlength{\itemsep}{\bibitemsep}%
      \setlength{\parsep}{\bibparsep}}%
      }
  {\endlist}
  {\item}

\DeclareNameAlias{sortname}{first-last}

\renewcommand{\bibfont}{\footnotesize}

\usepackage[nolist,withpage]{acronym}
\begin{acronym}
	\acro{CI}{Critical Infrastructure}
	\acro{LTE}{Long Term Evolution}
	\acro{MAC}{Media Access Control}
	\acro{RLC}{Radio Link Control}
	\acro{UE}{User Equipment}
	\acro{eNB}{evolved Node B}
	\acro{OFDM}{Orthogonal Frequency-Division Multiplexing}
	\acro{PUSCH}{Physical Uplink Shared Channel}
	\acro{PUCCH}{Physical Uplink Control Channel}
	\acro{PDCCH}{Physical Downlink Control Channel}
	\acro{RNTI}{Radio Network Temporary Identifier}
	\acro{MIMO}{Multiple Input Multiple Output}
	\acro{PSS}{Primary Synchronization Signal}
	\acro{PCI}{Physical Cell Identity}
	\acro{PCFICH}{Physical Control Format Indicator Channel}
	\acro{CFI}{Control Format Indicator}
	\acro{DCI}{Downlink Control Information}
	\acro{CRC}{Cyclic Redundancy Check}
	\acro{PDSCH}{Physical Downlink Shared Channel}
	\acro{HARQ}{Hybrid Automatic Repeat Request}
	\acro{ACK}{Acknowledgment}
	\acro{NAK}{Negative Acknowledgment}
	\acro{CQI}{Channel Quality Indicator}
	\acro{PRACH}{Physical Random Access Channel}
	\acro{DoS}{Denial of Service}
	\acro{IMSI}{International Mobile Subscriber Identity}
	\acro{SIM}{Subscriber Identity Module}
	\acro{SNR}{Signal-to-Noise Ratio}
	\acro{API}{Application Programming Interface}
	\acro{WAMPAC}{Wide Area Monitoring Protection and Control}
	\acro{IEC}{International Electrotechnical Committee}
	\acro{SV}{Sampled Values}
	\acro{SDR}{Software-Defined Radio}
	\acro{SISO}{Single Input Single Output}
	\acro{EPC}{Evolved Packet Core}
	\acro{MME}{Mobility Management Entity}
	\acro{S-GW}{Serving Gateway}
	\acro{P-GW}{Packet Gateway}
	\acro{IP}{Internet Protocol}
	\acro{UDP}{User Datagram Protocol}
	\acro{5G}{5th Generation of Cellular Mobile Communication}
	\acro{ICT}{Information and Communication Technology}
	\acro{RRC}{Radio Resource Control}
	\acro{OFDMA}{Orthogonal Frequency Division Multiple Access}
	\acro{CoMP}{Coordinated Multi Point}
	\acro{RAN}{Radio Access Network}
	\acro{PLMN}{Public Land Mobile Network}
	\acro{MIB}{Master Information Block}
	\acro{SIB}{System Information Block}
	\acro{CRS}{Cell Specific Reference Signal}
	\acro{DMRS}{Demodulation Reference Signal}
	\acro{RB}{Resource Block}
	\acro{SC-FDMA}{Single-Carrier Frequency Division Multiple Access}
	\acro{PRATTLE}{Physical Radio Access Termination in LTE}
	\acro{SRB}{Signalling Resource Bearer}
	\acro{DRB}{Dedicated Resource Bearer}
\end{acronym}

\renewcommand{\baselinestretch}{0.971}\normalsize

\begin{document}

\title{Towards Resilient 5G: Lessons Learned from Experimental Evaluations of LTE Uplink Jamming}
\author{
\IEEEauthorblockN{Felix Girke, Fabian Kurtz, Nils Dorsch, Christian Wietfeld\\
Communication Networks Institute\\
TU Dortmund University\\
\{felix.girke, fabian.kurtz, nils.dorsch, christian.wietfeld\}@tu-dortmund.de}
}

\maketitle

\IEEEaftertitletext{\vspace{-2\baselineskip}}

\notice

\begin{abstract}
	Energy, water, health, transportation and emergency services act as backbones for our society.
	Aiming at high degrees of efficiency, these systems are increasingly automated, depending on communication systems.
	However, this makes these \aclp{CI} prone to cyber attacks, resulting in data leaks, reduced performance or even total system failure.
	Beyond a survey of existing vulnerabilities, we provide an experimental evaluation of targeted uplink jamming against \ac{LTE}'s air interface.
	Primarily, our implementations of smart attacks on the \ac{LTE} \ac{PUCCH}, the \ac{PUSCH} as well as on the radio access procedure are outlined and tested.
	In exploiting the unencrypted resource assignment process, these attacks are able to target and jam specific UE resources, effectively denying uplink access.
	Evaluation results reveal the criticality of such attacks, severely destabilizing \aclp{CI}, while minimizing attacker exposure.
	Finally we derive possible mitigations and recommendations for \acs{5G} stakeholders, which serve to improve the robustness of mission critical communications and enable the design of resilient next generation mobile networks.
\end{abstract}

\acresetall

\IEEEpeerreviewmaketitle

\section{Introduction}
Nowadays, societies heavily depend on services provided by so-called \acp{CI}, including energy, water, health, transportation, public safety and communication systems.
To improve efficiency and management of such infrastructures, comprehensive automation is pursued, necessitating an integration of \ac{ICT}.
Due to the lengthy and costly deployment of dedicated wired networks, harnessing (public) mobile communication technologies is widely regarded a suitable approach.
Yet, ubiquitous connectivity eases access not only for authorized users, but also for malicious third parties.
Cyber attacks on \ac{CI} communication networks can severely degrade functionality and system stability.
An attack, e.g. aimed at the power grid's \ac{ICT} infrastructure, could trigger events leading to outages or blackouts \cite{smartgrid_security}.
Two major ways of disrupting a \ac{CI}'s wireless communication can be distinguished.
On one side, so-called barrage jammers may be employed, which essentially obscure all user signals in a certain frequency range by using wideband noise.
Such jammers are very effective in disrupting services and straightforward to implement.
Yet, they can be detected and located with low effort, allowing authorities to stop the attack and hold attackers accountable.
On the other side, smart jammers exploit inherent system properties such as protocol flaws, requiring less power.
Thus, highly precise, covert attacks are enabled, only affecting target devices instead of an entire area or frequency band.
Subsequently, attacks cannot be recognized as easily.
However, this class of jammers requires more advanced technological skills and knowledge.
Hence, smart jammers are of major interest for groups with sufficient resources and technical capacities, which want to perform attacks, while staying hidden.
Examples include hostile intelligence services or well-funded terrorist groups.
Employed efficiently, smart jammers provide the means to secretly undermine critical, public infrastructures which depend on wireless communications.

It has to be emphasized that this publication is not intended as guideline for such groups, but rather serves to indicate vulnerabilities and derive mitigation strategies.
Therefore, the main contributions of this paper can be summarized as below:
\begin{itemize}
	\item General overview of possible attack vectors against current cellular mobile communication networks,
	\item Design of a smart jammer for issuing attacks on the \ac{LTE} \ac{PUCCH} and \ac{PUSCH}, including corresponding evaluation results,
	\item Recommendations for improving \ac{CI} communication systems, focusing on \acs{5G} developments.
\end{itemize}
The remainder of this paper is structured as follows:
Section~\ref{sec:attacks} provides a survey of key jammer properties and known attacks.
It is followed by an overview of related work (Sec.~\ref{sec:relwork}).
Next, we introduce our smart uplink jammer (Sec.~\ref{sec:concept}) and present evaluation results (Sec.~\ref{sec:results}).
Recommendations for increased resiliency are given in Section~\ref{sec:recomm}.
Finally, Section~\ref{sec:conclusion} provides a summary and an outlook on future work.

\section{Jammer Properties, Possible Attacks\\and Mitigations}
\label{sec:attacks}
This section provides an overview of key jammer properties and possible attack vectors in \ac{LTE} infrastructures.

\subsection{Jammer Properties}
To disrupt radio communications, barrage jammers allow configuration of the used power level, bandwidth and duty cycle.
Higher power levels increase the affected area and boost the impact on user signals, yet make the attacker more vulnerable to detection.
Duty cycles indicate periods, in which the jammer is active.
In contrast to barrage jammers, which are limited to the above mentioned properties, smart jammers provide the following additional parameters:

\textbf{Synchronization}: Synchronizing jamming to target signals maximizes an attack's impact, while minimizing visibility.

\textbf{Eavesdropping}: For the same reasons, gathering information about a target system before initiating an attack is beneficial.
In \ac{LTE}, this task is facilitated by the lack of \ac{MAC} and \ac{RLC} encryption.
Thus, subscribers may be de-anonymized \cite{rupprecht-19-layer-two} or \ac{UE} locations can be tracked \cite{gutiguti}.
In case of stationary \acp{UE}, e.g. in Smart Grids, a high level of channel awareness could be obtained and utilized by the attacker.

\textbf{Placement}: Due to their small form factor, smart jammers allow for arbitrary placement.
Preferable locations may be close to the \ac{UE} for downlink or near the \ac{eNB} for uplink jamming.
Thus, lower power levels can be employed, minimizing attack visibility.
In contrast, high powered barrage jammers may be of significant size.

\textbf{Directionality}: Smart jammers may utilize highly directional antennas, focusing on specific targets. This increases range while decreasing power consumption, visibility and size.

\subsection{Attack Vectors}
\ac{LTE} \acp{RAN} can be attacked at the physical layer, physical channel and protocol level.
Examples of each category are given below, summarized by Figure \ref{fig:attack_vectors}.

\begin{figure*}[t]
	\centering
	\includegraphics[width=1\textwidth]{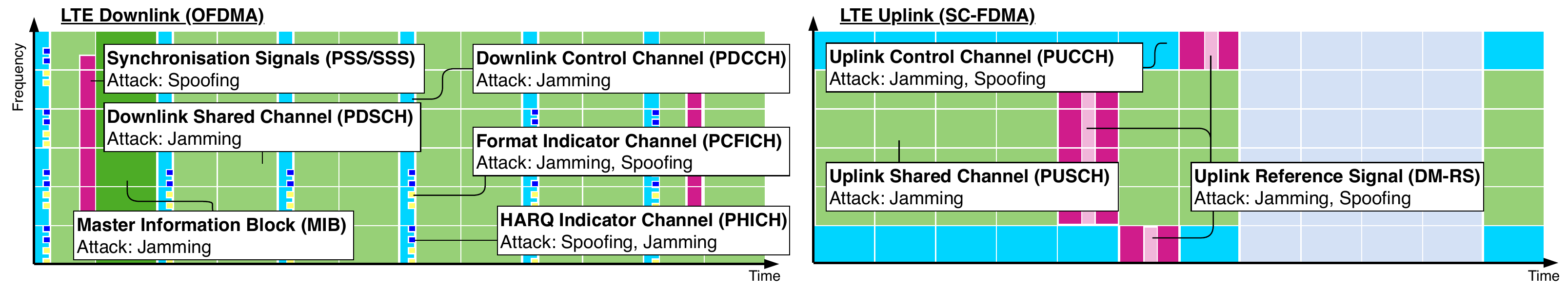}
	\caption{Overview of selected attack vectors, targeting the \ac{LTE} down- and uplink}
	\label{fig:attack_vectors}
	\vspace*{-5mm}
\end{figure*}

\subsubsection{Physical Layer Attacks}
\hfill\\
\textbf{Cyclic Prefix Attack}: As a key component of the \ac{OFDM} signal, the cyclic prefix prevents inter symbol interference and enables frequency domain equalization.
To degrade radio performance, noise can be emitted during cyclic prefix transmission.
Such an attack is considered highly effective as no specific mitigation is known, but it requires near perfect knowledge of the channel.

\textbf{Synchronization Attack}: In the course of a synchronization attack, the \ac{UE} is prevented from receiving the \ac{PSS}/\ac{DMRS} (down-/uplink) by shifting the symbol timing peak.
Here, the main challenge is locking onto the target signal.

\textbf{Reference Signal Attack}: Reference signal (pilot) attacks (\ac{CRS}/\ac{DMRS}) cause faulty channel estimations.
Thereby, diverging conditions are assumed, limiting physical channel throughput.
As downlink pilot signals are sparse, an attack requires a minimal duty cycle and transmission power.
However, precise synchronization based on an analysis of channel conditions is required.

\subsubsection{Physical Channel Attacks}
\hfill\\
\textbf{\ac{PCFICH} Jamming}: This method corrupts the channel format indicator.
First, the jammer synchronizes to the cell, to receive and decode the \ac{MIB}.
Next \ac{PCFICH} elements are calculated and noise or a modulated signal is transmitted on top of them.
If the corrupted \ac{CFI} is higher than the actual one, the \ac{UE} tries to decode non-existent \acp{DCI}.
In case the corrupted \ac{CFI} is lower, the \ac{UE} overlooks \acp{DCI}.
For both scenarios, the first subframe slot's \ac{PDSCH} symbols are expected in the wrong location, causing errors in the corresponding transport blocks.

\textbf{\ac{PDSCH} Jamming}: The \ac{PDSCH} carries configuration as well as downlink user data.
Hence, it allows targeting one or multiple \acp{UE}.
Precise synchronization and \ac{DCI} decoding are required to locate resources of a particular \ac{UE}.

\textbf{Paging Jamming}: Here \ac{PDSCH} jamming is modified, by targeting paging channel \acp{RB}.
The channel informs idle \acp{UE} of pending downlink data.
If respective \acp{RB} are jammed, the \ac{UE} is never notified of transmissions and remains idle.
This attack reduces a jammer's duty cycle to predictable paging periods, relaxing \ac{DCI} decoding requirements.

\textbf{\ac{PUSCH} Jamming}: The purest version of \ac{PUSCH} jamming is achieved by using a barrage of white noise on this channel.
It is straightforward to implement, enabling \ac{DoS} throughout an entire cell.

\textbf{\ac{PUCCH} Jamming}: The \ac{PUCCH} is located at the edge of the uplink bandwidth.
Thus, it can be identified and jammed with low effort, affecting many \acp{UE} at once.
The attack can lead to misinterpretation of received signals as scheduling requests, \ac{HARQ} (negative) acknowledgments, \acp{CQI} or \ac{MIMO} precoding matrices, drastically reducing up- and downlink throughput.
While some control data may be multiplexed into the \ac{PUSCH}, use of the \ac{PUCCH} enables higher cell capacities and data rates (e.g. via \ac{MIMO}).
Another advantage for the adversary is that the \ac{PUCCH} power limit is commonly lower than the \ac{PUSCH} power limit in order to reduce inter-cell interference.

\textbf{UE Targeted Uplink Jamming}: Since the entire uplink (both \ac{PUSCH} and \ac{PUCCH}) is slaved to the downlink with \SI{4}{ms} delay, attackers can identify and target a particular \ac{UE}'s \acp{RB}.
This is due to the nature of the \ac{PDSCH}, which relies on obfuscation via \ac{CRC} mask rather than proper encryption.
An implementation and evaluation of such a jammer is detailed in Section \ref{sec:concept}.
Given a \ac{DCI} decoder, a target device's \acp{RB} can be found via its \ac{RNTI}, obtainable by combining de-anonymization \cite{gutiguti} with tools like C3ACE \cite{Falkenberg2017c}.

\subsubsection{Protocol Attacks}
\hfill\\
\textbf{Cell Barring Spoofing}:
Rogue \acp{eNB} mimic real cells, barring \acp{UE} from connecting to specific cells by transmitting  \textit{cellBarred} and \textit{intraFreqReselection} flags.
Thus, \acp{UE} will ignore any cells on that frequency.
Also, \ac{LTE} requires \acp{UE} to ban cells from which \acp{SIB} or \acp{MIB} are not received within a defined time frame.
Hence, corrupting these blocks leads to barred cells for up to \SI{300}{s} as well \cite{3gpp_rel14_ts_36.304}.

\textbf{Reject Spoofing}:
An identical rogue cell is created by synchronizing to a legitimate cell and mimicking its relevant parameters such as \ac{PLMN} ID, tracking area, etc. 
Next, target \acp{UE} are forced from their original cell, and the rogue cell rejects subsequent reattachment attempts.
Depending on the specific implementation, \acp{UE} may treat the \ac{PLMN} as generally barred.
Thus, this attack can be very efficient, yet requires significant effort.

\section{Related Work}
\label{sec:relwork}
In recent years, several publications have dealt with the analysis of different \ac{LTE} vulnerabilities.
An overview of various attack vectors against the \ac{LTE} air interface is given in \cite{lichtman_overview}, estimating potential jammer to signal power ratios.
Hussain et al. propose LTEInspector \cite{lteinspector}, which combines a symbolic model checker with a cryptographic protocol verifier.
It is applied to analyze several different attacks, most of which are validated using a testbed set-up.
Though commercial \acp{UE} are used for verification, evaluation is mainly focused on protocol implementations and core network signaling.
Cyclic prefix and pilot jamming are evaluated for both up- and downlink in \cite{cp_pilot_jamming}.
Measurements indicate that \ac{OFDMA} is more resilient against these types of attack than \ac{SC-FDMA}.
A method for disrupting the \ac{PCFICH} is introduced in \cite{pcfich_jamming}, using simulations for validation.
However, real-world measurements would be required for a comprehensive assessment.
Also, \acs{5G} will mitigate the threat by removing the \ac{PCFICH}.
Labib et al. use radio frequency spoofing to impair synchronization and cell selection \cite{labib_cell_selection,labib_cell_sync}. %
Based on experiments with a software \ac{UE}, enhancements to the aforementioned techniques are proposed.
Simulations are performed in \cite{lichtman_pucch} to evaluate jamming against different \ac{PUCCH} formats. %
Possible mitigations are discussed, e.g. foregoing \ac{PUCCH} transmissions at the cost of limiting system capacity.
\ac{DoS} and de-anonymization (\ac{IMSI} catching) attacks are evaluated against commercial \acp{UE} in a laboratory by \cite{mjolsnes_imsi_catcher}.
Yet, the employed \ac{eNB} is not commercial grade.
Rao et al. provide physical layer measurements on \ac{LTE} resilience, considering attacks on different features of the \ac{OFDMA} downlink signal \cite{Rao2017LTEPL}.
However, the same open-source \ac{eNB} and \ac{UE} implementations are evaluated against each other.
Thus, results cannot be generalized to commercial equipment.\\
In comparison to related work, this publication evaluates further aspects of the \ac{LTE} air interface, considering the impact on end-to-end connectivity of commercial grade equipment.

\section{Targeted Uplink Jammering:\\Concept, Implementation and Evaluation Scenario}
\label{sec:concept}
Within this section, we introduce our approaches to jamming and describe the scenario and measurement setup.

\subsection{Smart Uplink Jamming Concepts}
\subsubsection{PUSCH/PUCCH Jammer}
Our jammer, implemented on top of srsLTE \cite{gomez2016srslte}, disturbs the \ac{LTE} \ac{PUSCH} and \ac{PUCCH} respectively.
To perform the attack, we synchronize the device with a cell, decode the \ac{PDCCH} and transmit on \acp{RB}, originally assigned to the target \ac{UE}.
Building on the \ac{UE} implementation in srsLTE, the jammer imitates the victim device.
Upon activation, the jammer jumps to the \ac{RRC} Connected State immediately, utilizing the provided target \ac{RNTI}.
Afterwards, the jammer is able to transmit on top of the \ac{UE}'s assigned \acp{RB}.
No other radio access or scheduling requests are issued and \ac{RRC} remains nearly unchanged.

\subsubsection{\ac{PRATTLE}}
\begin{figure}[t]
	\centering
	\includegraphics[width=1\columnwidth]{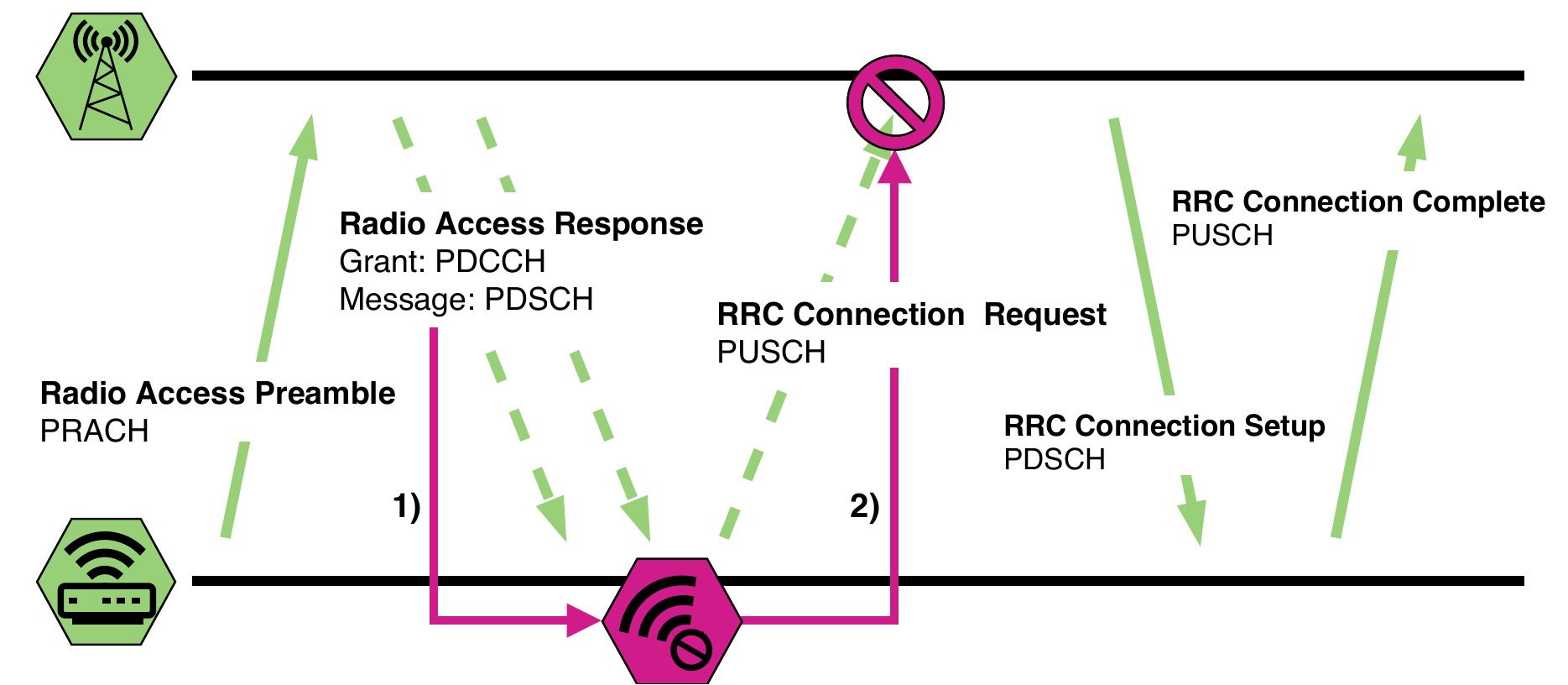}
	\caption{Principle of the PRATTLE attack}
	\label{fig:prattle}
	\vspace*{-5mm}
\end{figure}

Besides the above described jammers, an attack on the radio access procedure is devised, referred to as \ac{PRATTLE}.
It targets the first \ac{PUSCH} message of one or several \acp{UE}.
The attack is structured as shown in Figure \ref{fig:prattle}:
1) The jammer continuously monitors the \ac{PDCCH} for Radio Access Response Grants.
2) \ac{RRC} Connection Requests, identified by grants, are jammed repeatedly.
3) Due to this attack, the \ac{UE} fails the complete procedure and retries until a maximum number of attempts (6 to 200) is reached.
After that, the corresponding cell is to be treated as barred for up to \SI{300}{s} \cite{3gpp_rel14_ts_36.304}.
In the following, however, only the former two attacks are evaluated.

\subsubsection{Critical Infrastructure Jamming Scenario}
To evaluate the developed uplink jammer's effectiveness, we consider a Smart Grid scenario.
Such \acp{CI} possess well-defined requirements, as specified by e.g. the \ac{IEC}'s standard 61850 \cite{61850-5:TC57:IEC}, which are extremely challenging regarding reliability and latency.
Moreover, cyber attacks on power grid \ac{ICT} infrastructures may be of severe consequences, endangering all dependent systems.
Attackers profit from the static nature of grid assets, allowing for optimal target localization and jammer placement.
We specifically consider \ac{IEC}~61850 based \ac{WAMPAC} systems, regularly transmitting measurement data between Smart Grid substations.

\begin{figure}[t]
	\centering
	\includegraphics[width=1\columnwidth]{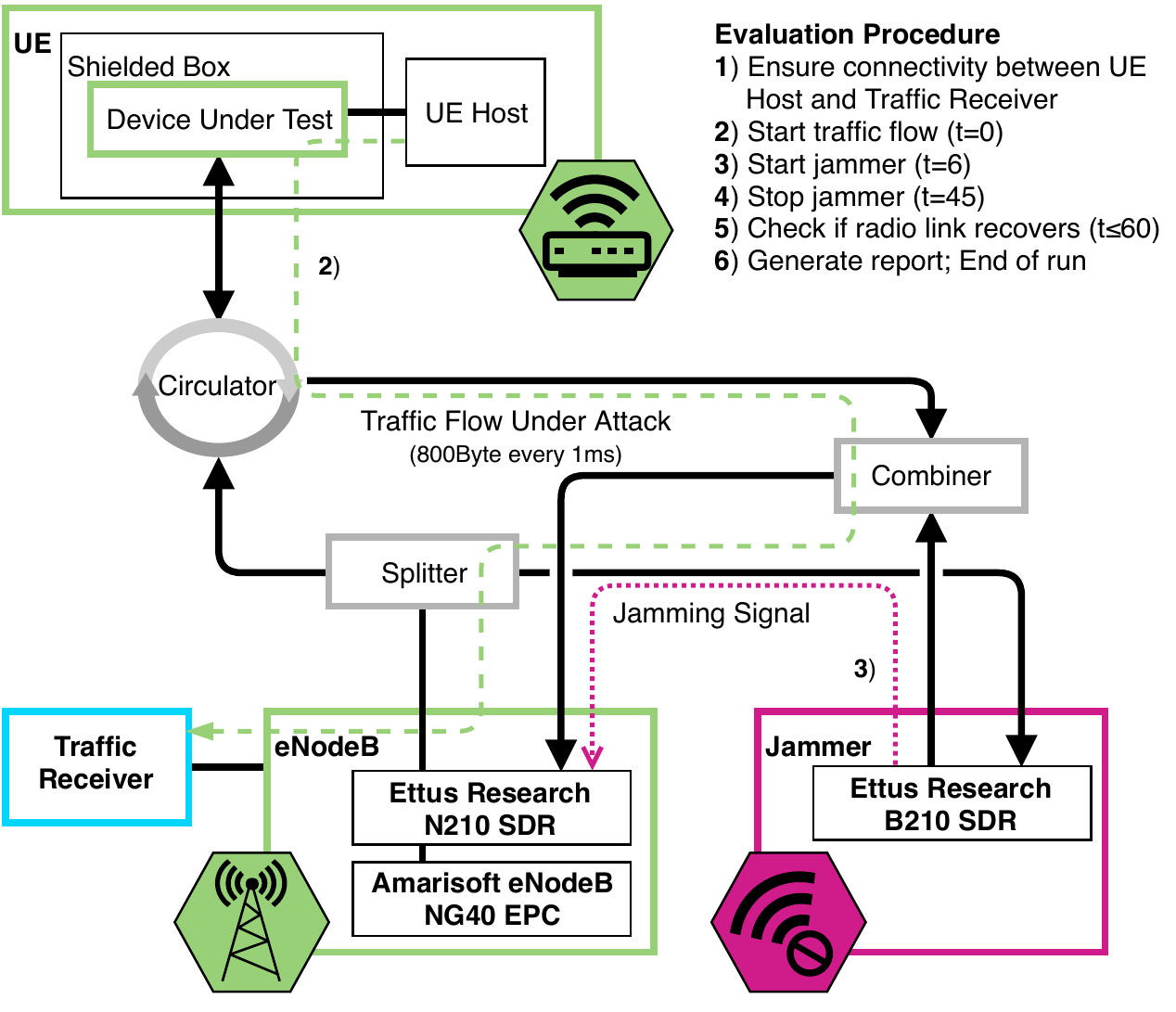}
	\caption{Experimental setup for evaluating the uplink jammer}
	\label{fig:setup}
	\vspace*{-5mm}
\end{figure}

\subsection{Experimental Measurement Setup}
Measurements are performed within our laboratory as shown in Figure \ref{fig:setup}, in isolation from public radio networks.
For this purpose, the \ac{UE} under test is placed in a shielding box and connected to the \ac{eNB} via a box-internal antenna.
\ac{UE} uplink and jammer signal are combined and fed to the \ac{eNB}.
For synchronization, the downlink signal of a commercial, \ac{SDR}-based Amarisoft base station is sent to \ac{UE} and jammer.
An Ettus Research N210 \ac{SISO} / 2x2 \ac{MIMO} \ac{SDR} serves as radio frequency frontend.
The \ac{eNB} supports dynamic power control, status displays, an \ac{API} for retrieving diagnostic data, information on active \acp{RNTI} and \ac{UE} identities.
It is configured with a bandwidth of \SI{20}{MHz} and parametrized to emulate an observed real world cell.
Transmit and receive gain of the \ac{eNB} are adjusted to achieve stable, near optimal radio link conditions.
\ac{MME}, \ac{S-GW} and \ac{P-GW} functionalities are provided by an NG40 Virtual \ac{EPC}.
The jammer itself utilizes the Ettus Research B210 \ac{SDR} platform, which is recommended for use with srsLTE.
As victim \acp{UE}, we employ a Sierra Wireless Air Prime EM7455 (Qualcomm Snapdragon X7 LTE Modem, Cat. 6) and a Huawei ME909s-120 (HiSilicon Balong 711, Cat. 4), as both devices enable an automated evaluation process.
Additional qualitative evaluations with an LG G5 smartphone (Qualcomm X12 LTE) confirm the results.
Measurements are conducted for \ac{PUSCH} and \ac{PUCCH} at different jammer gain levels, with two \ac{LTE} implementations (i.e. \acp{UE}).
The transmission of measurement values according to \ac{IEC}~61850, is replicated with our purpose-built traffic generator, offering a higher degree of flexibility than standard tools.
Initial experiments show that the \ac{eNB} limits the \ac{UE}'s capability of mitigating jamming-induced, degraded \ac{SNR}, as the \ac{eNB} aims to reduce the \ac{UE}'s transmission power for minimizing interference between users.
Further, our evaluations indicate high sensitivity towards variations of \ac{LTE} set-up parameters.

\section{Evaluation Results}
\label{sec:results}
For evaluation, measurement values are transmitted every \SI{1}{ms} via \SI{800}{Byte} \ac{UDP} packets.
Jammer gain is increased from \SI{1} to \SI{35}{dB} with a step size of \SI{2}{dB}.
At every gain level $20$ runs of \SI{60}{s} duration each are performed.
The jammer is active for \SI{39}{s} between $t_{start}=\SI{6}{s}$ and $t_{end}=\SI{45}{s}$.
We employ the following measurement procedure: 1) start of \ac{eNB}, \ac{UE} and traffic generation/reception servers, 2) initiation of the \ac{UE}'s packet stream, 3) jammer start, 4) jammer stop, 5) recovery period of the \ac{UE}-\ac{eNB} link, 6) waiting for the traffic generator's termination packet (if the link is re-established) and the receiver's report.

\subsubsection{PUCCH vs. PUSCH jamming}
\begin{figure}[b]
	\vspace*{-4mm}
	\centering
	\includegraphics[width=1\columnwidth]{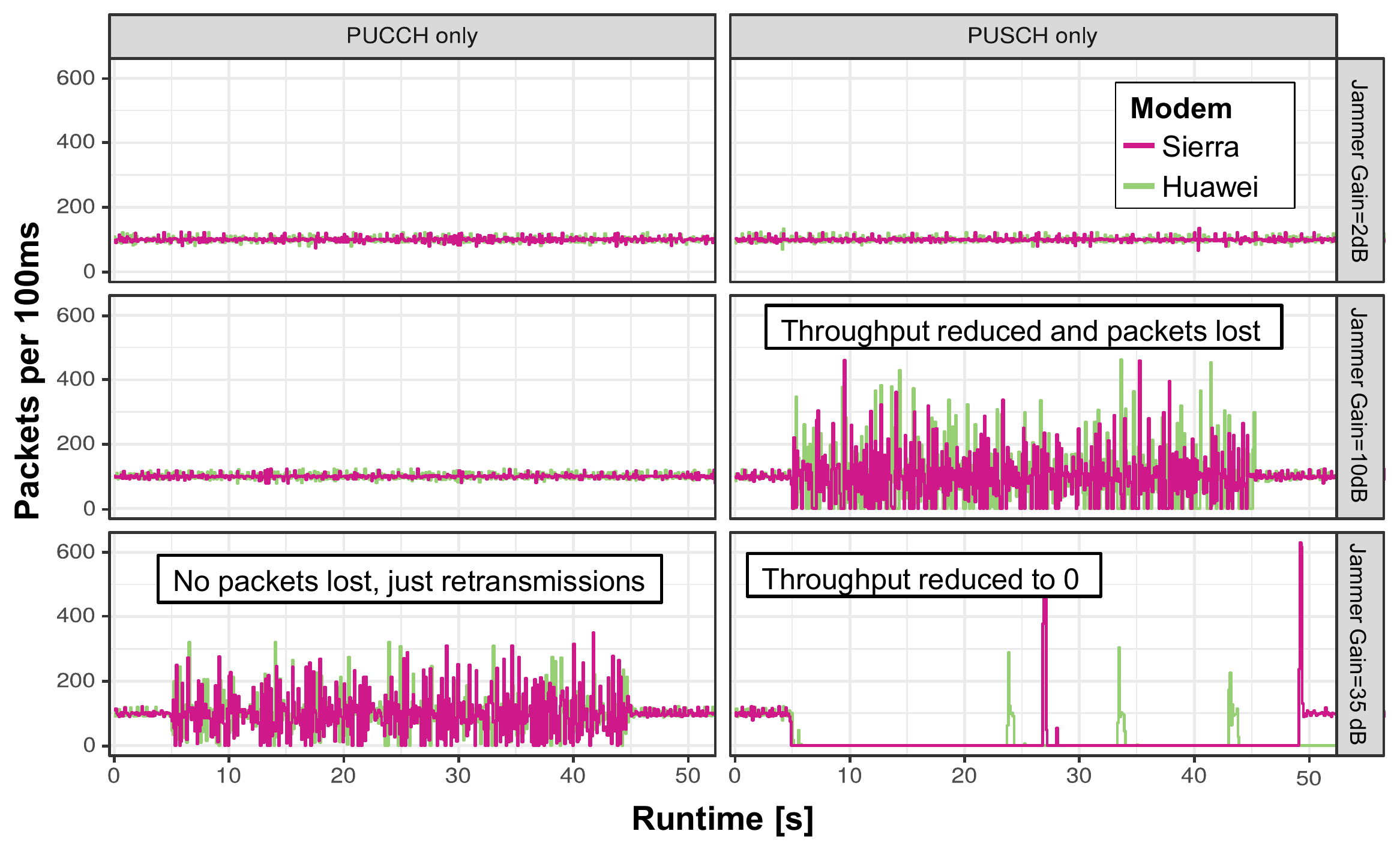}
	\caption{Comparison of the number of received packets under the impact of \ac{PUCCH} and \ac{PUSCH} jamming, considering three jammer gain levels for two commercial modems}
	\label{fig:pusch_pucch}
\end{figure}

Figure \ref{fig:pusch_pucch} gives exemplary results of both tested \acp{UE} at three different jammer gain levels.
Repeated runs show similar behavior.
For low gains (about $\SI{2}{dB}$), it can be noted that \ac{PUCCH} jamming does not affect the connection.
Applying high gains, retransmissions occur without loss of packets.
However, for \SI{10}{dB} jamming retransmissions occur, reducing throughput.
Packets are dropped as the \ac{UE} is unable to drain its transmit buffer.
Boosting gain further reduces throughput to zero.
\ac{PUCCH}-only jamming has a low impact as the \ac{eNB} shifts control data to the \ac{PUSCH}.

\subsubsection{PUSCH jamming for different UE implementations}
\begin{figure}[t]
	\centering
	\includegraphics[width=1\columnwidth]{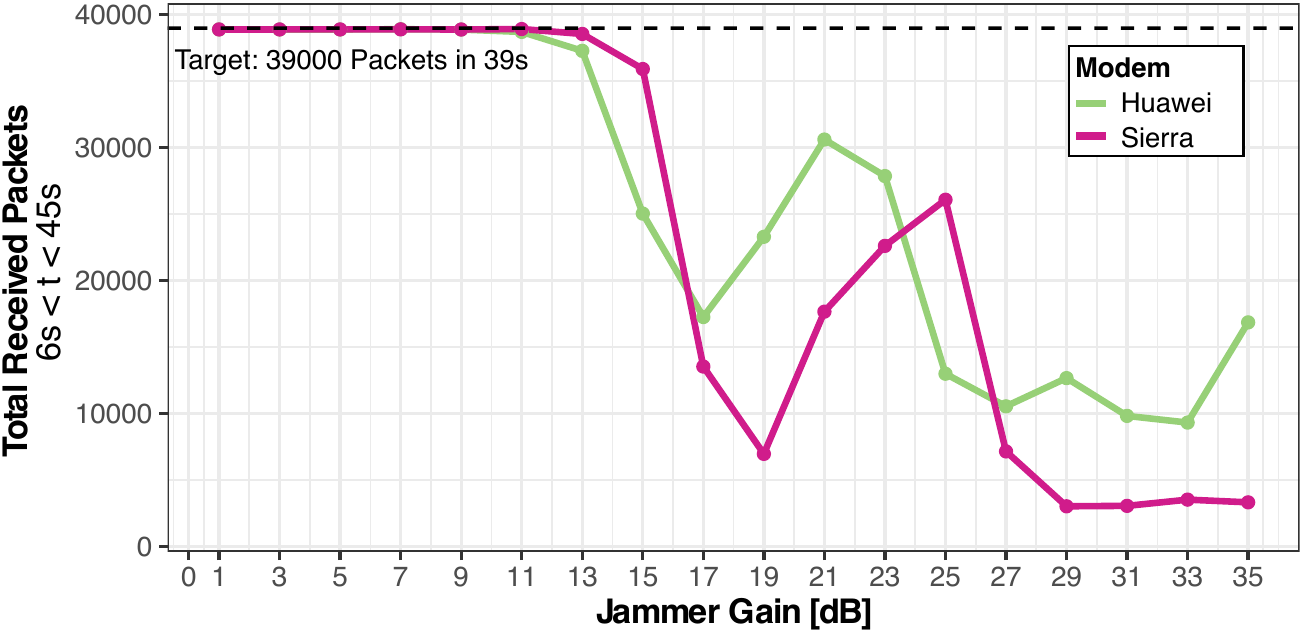}
	\caption{\ac{PUSCH} jammer impact under different jammer gains}
	\label{fig:pusch_overview}
	\vspace*{-5mm}
\end{figure}

The behavior of different \acp{UE} under \ac{PUSCH} jamming is compared in Figure \ref{fig:pusch_overview}.
It illustrates the total number of successfully received packets for different jammer gains.
Both \acp{UE} start to lose packets from gains of approximately \SI{11}{dB}.
Yet, at higher gain levels the two devices recover slightly, with the Huawei increasing throughput between \SI{19}-\SI{23}{dB} and the Sierra Wireless in the range of \SI{21}-\SI{25}{dB}.
This is caused by the \ac{eNB} detecting the worsening channel conditions.
It therefore grants additional resources and requests the \ac{UE} to use more robust modulation and coding schemes.
Furthermore, the Huawei \ac{UE} recovers at a gain of \SI{35}{dB}, which is explained in more detail in the following paragraph.
Overall the Huawei modem shows better resilience, achieving higher throughput for most configurations.
However, in contrast to the Sierra Wireless \ac{UE}, it frequently crashes, requiring manual restarts.

\begin{figure}[t]
	\centering
	\includegraphics[width=1\columnwidth]{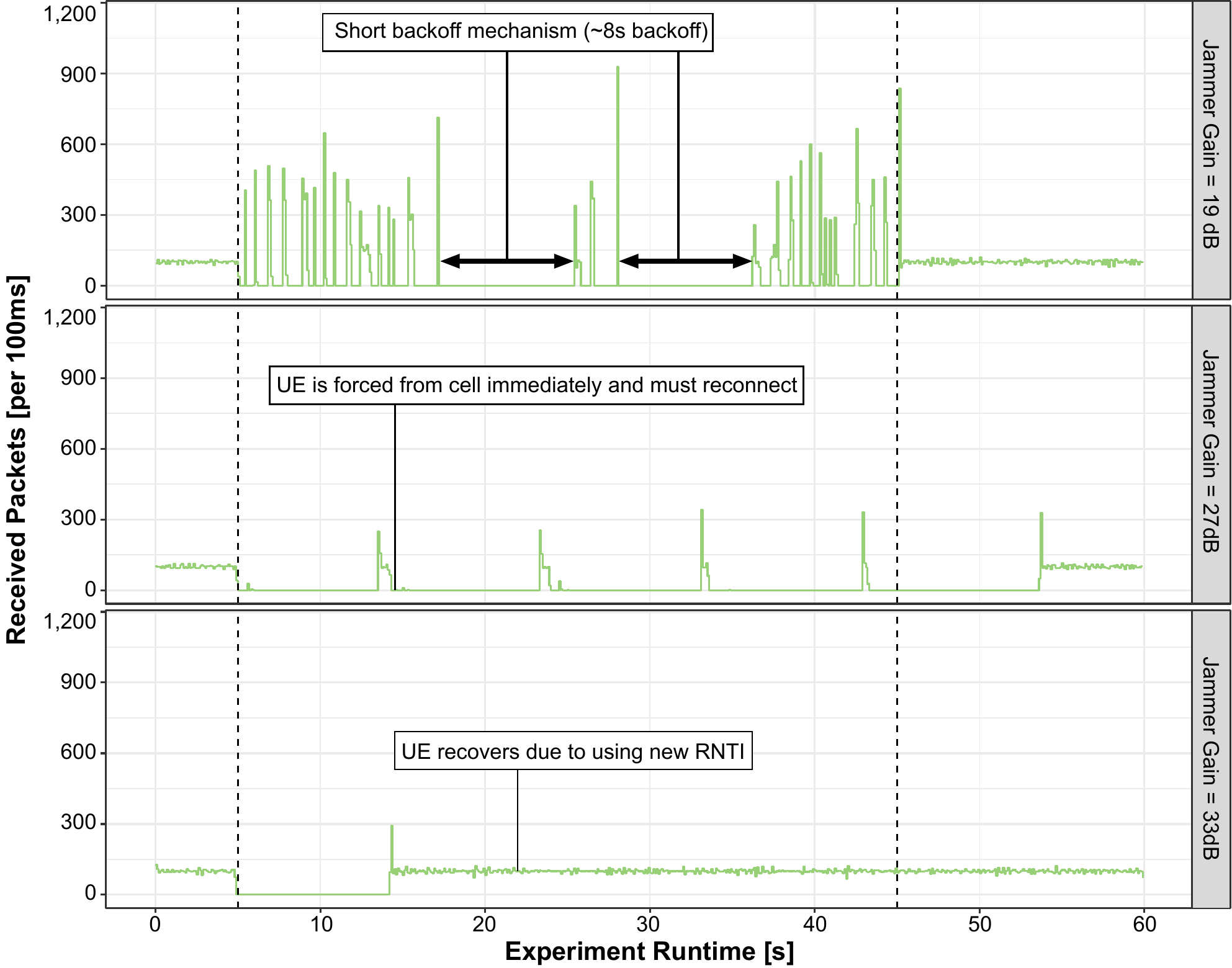}
	\caption{Detailed \ac{UE} behavior under \ac{PUSCH} jamming}
	\label{fig:pusch_details}
	\vspace*{-5mm}
\end{figure}

The following analysis provides additional details on the Huawei \ac{UE}'s reaction to \ac{PUSCH} jamming.
Figure \ref{fig:pusch_details} shows its uplink throughput over time for selected gains.
Again, runs with the same parameter set exhibit similar behavior.
The top part of Figure \ref{fig:pusch_details} (gain: \SI{19}{dB}) shows strongly fluctuating throughput starting at \SI{6}{s}, due to jamming induced link degradation.
At \SI{18}{s} (\SI{28}{s} respectively) the \ac{UE} loses \ac{eNB} connectivity and attempts to reconnect using its previous \ac{RNTI}.
Thus, the jammer is able to continue its attack as soon as the \ac{UE} reconnects.
After failing the radio access procedure several times, the \ac{UE} enters a back off period of (in this case) \SI{8}{s}.
With higher gains (middle of Figure \ref{fig:pusch_details}), the \ac{UE} is forced from the cell immediately after radio access ($\ll\SI{1}{s}$), as its signal is significantly weaker than the jammer's.
The bottom part of Figure \ref{fig:pusch_details} is specific to the Huawei device.
At a very high jammer gain of \SI{33}{dB} the connection is disrupted in a way, which causes the \ac{UE} to assume a complete connection loss.
Hence, it starts an entirely new connection, utilizing a different \ac{RNTI} and sending an \ac{RRC} Connection Request instead of an \ac{RRC} Reestablish Request.
In this way the \ac{UE} shakes off the jammer, achieving stable transmission after \SI{14}{s}.

Repeating experiments with the Sierra Wireless \ac{UE} show behavior similar to the Huawei modem.
However, the Sierra Wireless handset does not recover in any case.
Also, the \ac{UE}'s back off period increases with the number of failures (up to: Sierra \SI{22}{s}, Huawei \SI{12}{s}), i.e. at higher jammer gains.
In several cases it even refuses to connect for up to \SI{300}{s}.
This indicates that the cell barring timer is used to exclude the \ac{eNB} from its list of selectable cells.

\subsection{Uplink Jamming Mitigations}
An uplink jamming attack may be identified by the \ac{eNB} on basis of up- and downlink channel conditions diverging significantly from each other as well as from historical data.
Our evaluations using the Huawei device already point to a possible mitigation strategy.
It was shown that switching the \ac{RNTI} allows to render the attack ineffective, requiring the attacker to re-identify the \ac{UE}.
Hence, new, unpredictable \acp{RNTI} should be assigned on both connection and reconnection attempts.
Yet, since control channels in LTE are not encrypted, this is not a sustainable option.
Beyond that, \acp{eNB} should assign more robust modulation and coding schemes more quickly and allow critical \acp{UE} a higher transmit power.

\section{Recommendations for\\5G Standardization and Deployment}
\label{sec:recomm}
In the following we provide recommendations for 5G stakeholders organized by standardization and deployment aspects.

\subsection{Standardization of 5G Mobile Networks}
As a major aspect of \ac{5G} new radio, beamforming increases resilience against jamming.
However, there are still several shortcomings, which should be addressed by future \acs{5G} releases.
Though \acs{5G} finally supports encryption and integrity protection of \ac{SRB}$>0$ and \ac{DRB} \cite{3gpp_rel15_ts_38.331,3gpp_rel15_ts_33.501}, null encryption is still acceptable \cite{3gpp_rel15_ts_33.501}.
Even if encryption is applied, the air interface is not secured until after the \ac{UE} successfully completes the entire attach procedure \cite{3gpp_rel15_ts_33.501}.

\vspace{0.75mm}
\subsubsection{Mitigating Cell Barring and Reject Spoofing Attacks}
According to the \acs{5G} \ac{RRC} standard \cite{3gpp_rel15_ts_38.331} intra-frequency barring is still applied without verifying broadcasted barring information.
Such cell impersonation attacks could be prevented through the use of verification schemes for cell configuration (either for all \acp{UE} or those required by \ac{CI}).
Therefore, the following mechanism may be employed, using signatures applicable to both \ac{LTE} and \acs{5G}:
1) Operators generate public/private key pairs of reasonable size (e.g. \SI{2048}{bit}).
2) Operators provision public keys in the \acp{SIM}.
3) Operators sign hashes of concatenated \ac{MIB} and scheduled \ac{SIB} contents with their private keys
4) Base stations broadcast the signature in dedicated \acp{SIB}.
5) \acp{UE} receive all periodic \acp{SIB}, tentatively applying their settings, then verify the signature using the public key.
6) If verification succeeds, settings are committed (including intra-frequency cell barring), otherwise the settings are ignored and the cell is treated as malicious.
While this approach prolongs the cell search, it precludes barring spoofing and reject attacks.

\vspace{0.75mm}
\subsubsection{Mitigating Uplink Jamming Attacks}
As in \ac{LTE}, the 5G \ac{PDCCH} is not encrypted.
Hence findings still apply (with few limitations).
A simple means of increasing the computational load for the attacker is scrambling the entire \ac{DCI} with a sequence derived from the destination \ac{RNTI}, rather than just scrambling the \ac{CRC} (as in \ac{LTE}).
To compensate for limited \ac{DCI} entropy, an \ac{RNTI} hopping scheme can be employed at \acs{5G} base stations by periodically reconfiguring the \ac{RNTI}, once encryption for \ac{SRB}1 \cite{3gpp_rel15_ts_38.331} is established.
Thus, the probability of an attacker learning the \ac{RNTI} during its lifetime is reduced.\\
\indent Besides, some general measures can be used to improve the resilience of \acs{5G}.
Providing the standard in a machine readable formate would facilitate (automatic) identification of critical flaws.
Lightweight, mutual authentication and encryption for the air interface should be introduced.
Open, less complex, single-purpose protocols, such as Internet protocols, should be preferred to highly flexible solutions with many optional, potentially vulnerable extensions.
Also, legacy compatibility with insecure mechanisms and protocols is to be avoided.

\subsection{5G System Operation for Critical Infrastructures}
To identify irregularities in system operation, 5G mobile network operators should deploy advanced monitoring systems.
Rogue and impersonated \acp{eNB} can be identified with this method.
Beamforming antennas allow to increase \ac{SNR}, suppress interference signals and restrict jammer placement options, thus enhancing resilience.
Distributed beamforming techniques such as \ac{CoMP} provide more robust radio links by receiving signals from multiple \acp{eNB}.
Secure reallocation schemes for temporary \ac{IMSI} mitigate de-anonymization.
Disallowed or limited null-ciphering protects user data and improves overall security.

Prior to deployment, \ac{CI} operators ought to check coverage to ensure \acp{UE} are offered high signal strength connections via more than one \ac{eNB}, providing mobile network redundancy.
Location specific connection properties should be considered during \ac{UE} installation, e.g. directed antennas may be employed under line-of-sight conditions.
Operators should procure \acp{UE}, which support higher \ac{MIMO} schemes for spatial redundancy.
Locking \acp{SIM} into tracking areas mitigates rogue \acp{eNB} with different area codes.
Also, \ac{SIM} cards without voice services prevent de-anonymization through call-repetition and downgrade attacks, i.e. forcing \acp{UE} into insecure and slow 2G/3G cells.
\ac{CI} utilities and mobile network providers can improve attack detection by continuously exchanging performance data, e.g. via secure \acp{API}.
Also, appropriate layer~4 protocols help reduce the impact of jamming.

\vspace{1mm}

\section{Conclusion and Outlook}
\label{sec:conclusion}
Within this paper we provide an overview of relevant cyber attacks on mobile communication networks, endangering the stability of \acp{CI}.
In particular, we present our concept for disrupting uplink communications on the \ac{LTE} \ac{PUSCH} of a Smart Grid \ac{ICT} infrastructure.
Corresponding experimental evaluations demonstrate the attack's effectiveness.
The jammer is shown to significantly reduce throughput, forcing the \ac{UE} to back off or even crash.
Moreover, our analysis also indicates potential mitigations such as the application of scrambled or encrypted \acp{DCI}.
Recommendations to mobile network and \ac{CI} operators as well as to standardization bodies are derived, revealing a path towards secure and resilient \acs{5G}.
Key improvements involve beamforming and massive \ac{MIMO}.\\
In future work, we aim at broadening the evaluation to further equipment.
Also, the public radio access termination attack, described in Section \ref{sec:attacks}, is to be implemented and analyzed in detail.
For mitigation, we plan to develop a robust, lightweight \ac{DCI} encryption/scrambling strategy to counter eavesdropping, which serves as basis for attacks such as those presented here.

\section*{Acknowledgement}
\small
This work was supported by the federal state of Northrhine-Westphalia and the “European Regional Development Fund” (EFRE) 2014-2020 via the “CPS.HUB/NRW” project under grant number EFRE-0400008, and has been carried out within the Franco-German Project \textit{BERCOM} (FKZ: 13N13741), co-funded by the German Federal Ministry of Education and Research (BMBF).
The authors thank Stefan Monhof and Robert Falkenberg for valuable discussions.

\printbibliography

\end{document}